 \documentclass[final,5p,times,twocolumn]{elsarticle}

\usepackage{amssymb}
\usepackage{amsmath}
\usepackage[dvipsnames]{xcolor}
\usepackage{graphicx}
\usepackage{dcolumn}
\usepackage{bm}
\usepackage[colorlinks]{hyperref}     
\usepackage[normalem]{ulem}
\newtheorem{assumption}{Assumption}
\usepackage{ulem}
\usepackage[utf8]{inputenc}

\usepackage{tikz} 
\usetikzlibrary{calc,trees,positioning,arrows,chains,shapes.geometric,%
	decorations.pathreplacing,decorations.pathmorphing,shapes,%
	matrix,shapes.symbols}
\usetikzlibrary{calc,angles,positioning,intersections}

\usepackage[english]{babel}
\usepackage{subcaption}
\usepackage{comment}

\journal{Physics Letters B}
\begin{document}
\begin{frontmatter}


\title{Generalized relative locality and causal sets}

\author[first]{Andrea Bevilacqua\corref{cor1}}
\ead{andrea.bevilacqua@ncbj.gov.pl}

\author[first]{Alice Boldrin}
\ead{alice.boldrin@ncbj.gov.pl}

\cortext[cor1]{Corresponding author.}

\affiliation[first]{organization={National Centre for Nuclear Research},
            addressline={ul. Pasteura 7}, 
            city={Warsaw},
            postcode={02-093}, 
            country={Poland}}

\begin{abstract}

In this paper we introduce a new general framework for the study of phenomenological quantum gravity theories (PQG). The key idea is the introduction of two different types of spacetime, an observer-independent spacetime (modeled by a smooth orientable manifold) and an observer-dependent one (which has an inherently discrete causal set structure). The interaction between the two allows us to prove the main result of the paper: relative locality can be obtained in general PQG models, regardless of momentum-space curvature.  {We also discuss the treatment of spacetime symmetries in our model, and introduce a direct link between spacetime symmetries and relative locality effects.} Our construction is presented in a coordinate-independent way and is based on fibre bundles where spacetime, rather than momentum-space, is the base manifold. This makes causality manifest even in general models with relative locality. Furthermore, it allows for the application of this formalism to cosmology on general backgrounds, something which is not clearly possible in the canonical approach to relative locality, where momentum-space serves as the base manifold.

\end{abstract}

\begin{keyword}
Quantum Gravity \sep Relative Locality \sep Fibre Bundles \sep Causal Sets



\end{keyword}

\end{frontmatter}

\section{Introduction}

In his Tractatus Logico-Philosophicus, Wittgenstein famously argued that tautologies have nothing to say about the world \cite{wittgenstein1922tractatus}. However, while a tautology indeed relates two statements with the same logical content, the history of physics and mathematics is filled with examples of \textit{tautological} reformulations of known concepts in a new language which opens the way to further discoveries. The most famous example in physics is the reformulation of Newtonian mechanics in terms of Lagrangian or Hamiltonian.
This type of approach acquires an even greater importance in the context of quantum gravity (QG), where experimental guidance is at best sparse and theoretical or mathematical arguments are the main tools available to physicists. 

In this work, we consider the currently most promising approaches to QG research, particularly in terms of contact with experiments. We refer to these as phenomenological QG (PQG) models, in which additional parameters are introduced to describe potential QG signatures \cite{Addazi:2021xuf}. 
One of the most striking consequences of PQG models with a curved momentum-space is relative locality \cite{Amelino-Camelia:2011lvm, Freidel:2011mt, Amelino-Camelia:2011hjg}. The most natural way to model this class of theories is to assume that spacetime loses its fundamental role in the description of events, being replaced by momentum-space. Although conceptually appealing, this approach makes it challenging to describe relative-locality effects in general spacetimes, with existing discussions limited to highly symmetric or otherwise constrained cases \cite{Amelino-Camelia:2023srg, Fabiano:2025dsg}. Furthermore, causality is not manifestly well defined \cite{Chen:2012fu, Banburski:2013wxa, Amelino-Camelia:2014qaa}. Finally, while curvature of momentum-space is a sufficient condition for relative locality, it remains unclear whether it is also necessary.

In what follows we will propose an approach to relative locality which is based on the fibre bundle paradigm while still retaining the fundamental role of spacetime. One of the key concepts will be the separation between an underlying \textit{observer-independent} spacetime, and an \textit{observer-dependent} spacetime. In doing so we will provide a solution to all the problems delineated above, while at the same time introducing potentially new types of relative locality, for example related to the topology of momentum-space and/or spacetime. We will begin by introducing observer-(in)dependent spacetimes (sec. \ref{Secobservers}), further characterized in sec. \ref{connection} where dynamics will also be discussed. In sec. \ref{generalrelloc} we will show how the assumptions described in sec. \ref{Secobservers}, \ref{connection}, applied in the most simple context, reproduce literature results about relative locality. Before concluding in sec. \ref{conclusion}, we will discuss the curvature of momentum-space in sec. \ref{relloc} {, and introduce the treatment of symmetries in our formalism in sec. \ref{symmetries}}.

\section{Observer-independent and observer-dependent spacetimes} \label{Secobservers}

We consider a fibre bundle $(E,\pi,M)$ with fibres $F_p = \text{preim}_\pi(p)$, $p \in M$. We define the smooth orientable manifold $M$ to be the \textit{observer-independent} spacetime, with the fibres $F$ identified with momentum-space. Notice that the observer-independent spacetime lacks any properties that could lead to relative locality or causal sets. At the same time, we are not assuming anything about $F$ at this point, except that it is a smooth manifold. We now turn to the definition of observer-dependent spacetime.

\begin{figure}[t!]
\centering
\includegraphics[width=0.4\textwidth]{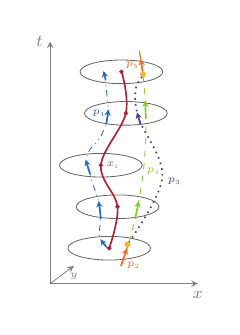} 
\caption{An observer moving along their worldline (solid red line). At each point of the trajectory, the observer can measure the momentum of particles that may enter or leave their observational scope  {(i.e. the balls $B_\epsilon$, represented in perspective as ellipses)}.  {The yellow dots represent interaction vertices between the orange, dashed green, and dotted purple particles.}}
\label{observers}
\end{figure}

\begin{figure}[t!]
\centering
\includegraphics[width=0.4\textwidth]{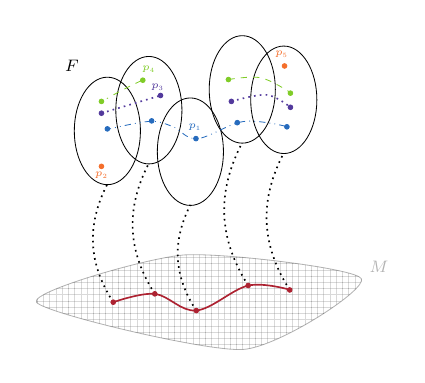} 
\caption{ {A representation of the same situation depicted in Fig. \ref{observers} but from the point of view of fibre bundles. Notice that $p_2$ and $p_5$, which are represented as dots here, are respectively the ending and starting point of the sections related to the incoming and outgoing orange particle.}}

\label{observers2}
\end{figure}

An observer moving along the timelike curve $x : [0,1] \to U \subset M$, $t \mapsto x(t)$, 
experiences spacetime locally through local sections $\sigma_{\theta(t)} : U_t \subset M \to E$. 
Here $U_t = B_\epsilon(x(t))$ denotes an open ball of radius $\epsilon$ centered at $x(t)$, 
and $\theta(t)$ is an index\footnote{The value of $\theta(t)$ may vary with the parameter $t$ along the curve $x(t)$.} 
labeling the particles visible to the observer along their trajectory. 
This reflects the idea that, in special and general relativity as well as in quantum field theory, an observer can only measure the four-momentum\footnote{In this paper we only focus on kinematical effects \cite{Amelino-Camelia:2011lvm, Freidel:2011mt}. We postpone the study of the measurement of the spin to future papers.} $p_\mu$ of particles within an arbitrarily small neighborhood of their worldline, while still being able to distinguish separate particle arrivals, a fact the observer interprets as the passage of time. Observers then reconstruct spacetime based on these measurements. Analogously, one can obtain the same result by identifying each spacetime point with an event. In the simplest case  {of a single freely propagating particle,} this corresponds to a 1-1 interaction: one particle arrives at the point with momentum $p$ and one particle leaves with momentum $q$ \cite{Freidel:2011mt}. By conservation of momentum we have $p = q$, so each spacetime point can be associated with a unique momentum.  {In more complex cases, where more general $n-m$ interactions are present, the observer assigns to the interaction point the set of all involved particles' momenta, reducing to the previous 1-1 case away from the interaction.} {In all cases, with or without interacting particles,} an observer may label nearby spacetime points by assigning a momentum to each visible particle, i.e. by choosing a corresponding  {set of} local section {s} of the bundle. The situation is depicted in Fig. \ref{observers}, \ref{observers2}.

The full description of spacetime is, however, not only given by the observation of particles' momenta at any given time, but also by the time evolution of the particle content, i.e., the temporal changes in particle number and types. We therefore define the following binary relation:
\begin{equation}\label{eqrel}
    \sigma_{\theta(t)}
    \sim 
    \sigma_{\theta(t')}
    \quad
    \text{iff}
    \quad
    \theta(s) = \theta(t) = \theta(t')
    \quad
    \forall s \in [t, t']_\pm
\end{equation}
where $[t, t']_\pm = [t, t']$ if $t\leq t'$, and $[t',t]$ otherwise. 
Relation \eqref{eqrel} is reflexive (since $[t,t] = \{t\}$), symmetric (if $\forall s \in [t,t']_\pm$, then $\forall s \in [t',t]_\pm$), and transitive (if $\forall s \in [t,t']_\pm$ and $\forall s \in [t',t'']_\pm$, then $\forall s \in [t,t'']_\pm$), and the equality $=$ is an equivalence relation. Hence, $\sim$ is an equivalence relation. We can now define the \textit{observer-dependent} spacetime as the set\footnote{One may wish to exclude elements in $S$ which describe zero particles seen by the observer. This amounts to a trivial modification of Eq. \eqref{oist}.} (see also Fig. \ref{observerdepst})
\begin{equation}\label{oist}
    S
    =
    \frac{\{\sigma^i_{\theta(t)}\}_{i, \theta(t)}}{\sim}.
\end{equation}
where $i$ runs over all observers.
Eq. \eqref{oist} expresses the idea that the observer-dependent spacetime is what all observers measure it to be. Because of this definition, when two observers interact at the same spacetime point $x \in M$, this event may still correspond to two distinct points in $S$. This reflects the fact that different observers perceive spacetime differently, even when meeting at the same event, a feature historically highlighted by the Rietdijk–Putnam argument \cite{Rietdijk_1966, 03588868-6265-3f83-975b-d9d64875974e}.

Following a similar line of reasoning, in \cite{Amelino-Camelia:2011lvm} observers are defined as `calorimeters with clocks'. However, that construction does not distinguish between an underlying observer-independent and an observer-dependent spacetime. Notice that here the observer-dependent spacetime $S$ has a purely phenomenological definition as we make no assumptions about the motion of particles in the vicinity of the observer. We only assume that the observer can measure the momentum of particles while moving along a trajectory $x(t)$ in $M$, and thus reconstruct the sections $\sigma_{\theta(t)}$. Note also that, in this model, an observer moves in the observer-independent spacetime $M$, which is, in general, different from the observer-dependent spacetime $S$ reconstructed from observations.
This seemingly troubling discrepancy is, in fact, already present in canonical special relativity, as exemplified by the Penrose–Terrell effect \cite{Terrell:1959zz, Penrose:1959vz, Hornof:2024rzt}. 
Moreover, in general relativity, different observers may adopt coordinate systems that do not cover the full spacetime, so that their observer-dependent description differs from the underlying observer-independent one. A well-known example is provided by the Schwarzschild spacetime, where standard Schwarzschild coordinates break down at the horizon \cite{Wald:1984rg}.
This situation presents itself in quantum mechanics too, where the measurement postulate introduces a clear distinction between the observer-independent unitary evolution of a system, governed by the Schrödinger equation, and the observer-dependent measurement process (collapse of the wave function).

\begin{assumption}\label{A1}
    \textit{We assume that the definition \eqref{oist} of observer-dependent spacetime $S$ holds in general PQGs. All physically observable quantities are to be described in terms of $S$.}
\end{assumption}

We also note that the observer-dependent spacetime $S$ has a discrete structure. The equivalence relation $\sim$ in Eq. \eqref{eqrel} identifies all the points in the trajectory of an observer in which there is no change in the particle content seen by the observer. Therefore, the discreteness of $S$ is ensured by the fact that the number of particles that an observer can see at any time is discrete, and so is the change in the particles observed. In other words, interestingly, the observer-dependent spacetime discreteness is ensured by the discrete nature of particles. 

 {
Note that our definition of observer-dependent spacetime $S$ in Eq. \eqref{oist} is a slight generalization of the canonical spacetime encountered in special and general relativity, where spacetime points are defined as events (intended as point-like interactions). The canonical specetime of special and general relativity corresponds to the limit $\epsilon \rightarrow 0$ of our construction (when the open balls $B_\epsilon$ become degenerate), so that observers are not able to measure the momentum of particles in a certain finite non-trivial nearby region, but only when they have an interaction vertex in common.\footnote{ {Our generalization accounts for the possibility of non-zero physical dimensions of observer apparatuses aimed at measuring particles momenta.}}
}

The final element needed to complete our framework is the connection on the bundle, related to the dynamics of the particles.

\section{Ehresmann connections and the causal set structure of $S$} \label{connection}

We start by briefly reviewing the concept of Ehresmann connection. Given a total space $E$ of a fibre bundle and its tangent space $TE$, one can split $TE = VE \oplus HE$ so that each vector in $TE$ is the sum of a \textit{vertical} component tangent to the fibres, and a \textit{horizontal} one complementary to it. As a general notation, given $w\in TE$, we will denote by $Vw$ and $Hw$ its vertical and horizontal components. A choice of smooth subbundle $HE$ is an Ehresmann connection.

In general, a system's (local) dynamics is described by some vector field in $TE$ defined through an action principle. Indeed, given a fibre bundle $(E,\pi,M)$ and calling $\mathcal{F}_U:= \Gamma^\infty(E,U \subset M)$ the space of smooth local sections $\sigma: U\subset M \rightarrow E$, one can define the equations of motion via the function $f: \mathcal{F}_U \rightarrow \mathcal{V}$, where $\mathcal{V}$ is some appropriate vector space. The field equations are then given by $f(\phi) = 0$ with $\phi \in \mathcal{F}_U$ \cite{Blohmann}. The set of solutions to the field equations is therefore $f^{-1}(0) \subset \mathcal{F}_U$, but instead of solving them explicitly, it is often useful to define $f^{-1}(0)$ as the set of the critical points of an action functional\footnote{In general, the action principle is not well defined, see \cite{Blohmann} for an extensive treatment of this issue.} $\mathcal{S}: \mathcal{F}_U \rightarrow \mathbb{R}$ \cite{Blohmann}. The critical points of $\mathcal{S}$ can equivalently be defined as integral curves of the vector field $w \in TE$ tangent to them.  Note that in general $w$ is not horizontal. A clear intuition for this is given by considering point-like particles, whose momentum can change along their trajectories implying a certain degree of motion along the fibres. Hence, a section describing the motion of a point-like particle is not in general horizontal. An alternative point of view is given by the determination of the dynamics through a Hamiltonian vector field $\xi_H$, defined in terms of the symplectic form $\omega$ by $i_{\xi_H}\omega = -dH$. In all cases, a choice of dynamics is equivalent to a choice of vector field in $TE$.

In general, PQGs include dimensional quantities $\lambda$, which, depending on the model, may parametrize either symmetry deformation or breaking \cite{Addazi:2021xuf}. For definiteness, in both cases we refer to such parameters as \textit{deformation parameters}. These considerations are included into our next assumption.

\begin{assumption}\label{A2}
    \textit{We assume that the local choice of dynamics amounts to a local choice of $w \in TE$ such that $w = v + \Lambda$, where $v \in TE$ describes the dynamics in absence of deformation, and $\Lambda$ is an Ehresmann connection which becomes degenerate as $\lambda \rightarrow 0$. 
    }
\end{assumption}

\begin{figure}[t!]
\centering
\includegraphics[width=0.5\textwidth]{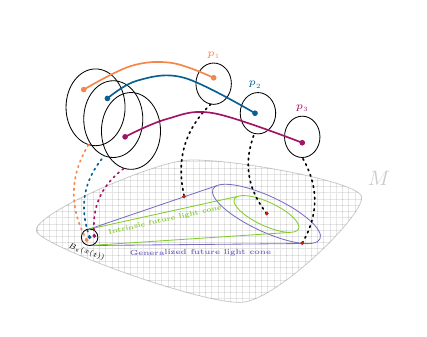} 
\caption{The generalized future light cone (outer one) of a ball $B_\epsilon(x(t))$ is the projection of integral curves of $w$, three of which are shown. The intrinsic light cone of $M$ (inner one), due to its Lorentzian structure, may not coincide with the generalized one.}
\label{lightcones}
\end{figure}

The choice of $v$ depends on the reference frame: different observers describe a system in different ways. Similarly, also the choice of $\Lambda$ is observer-dependent. This is in accordance, for instance, with \cite{Amelino-Camelia:2011lvm, Freidel:2011mt, Carmona:2021gbg}, where it was found that different observers observe locality differently.

Notice also that the identification of $f^{-1}(0) \subset \mathcal{F}_U$ through the action principle, common in special and general relativity, and quantum field theory, is not mandatory for (P)QG theories, or in general models which go beyond special/general relativity and quantum field theory. An example is given by the amplituhedron program \cite{Arkani-Hamed:2013jha}, where locality and unitarity (often explicitly introduced at the action level) are treated as emergent properties, and particles' behaviour is obtained in terms of a purely geometrical construct. The absence of any reference to an action principle in Assumption \ref{A2} is intended to allow the inclusion of these types of models in our framework. 

Assumption \ref{A2} naturally leads to assigning a causal set structure to the observer-dependent spacetime $S$, and to relative locality. In both cases, we require the base manifold to be additionally endowed with a Lorentzian metric. In this section we focus on the observer-dependent spacetime $S$, in sec. \ref{generalrelloc} we will discuss relative locality.

\begin{figure}[t!]
\centering
\includegraphics[width=0.4\textwidth]{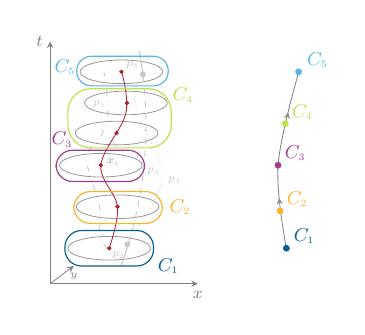} 
\caption{Left: An observer can see different particles along their motion. Here, $C_1 \rightarrow \theta = 4$, $C_2 \rightarrow \theta = 3$, $C_3 \rightarrow \theta = 1$, $C_4 \rightarrow \theta = 3$, $C_5 \rightarrow \theta = 4$.  {Notice that $C_1$ and $C_5$ include all the momenta of the particles involved in the interaction vertex. Note also that $C_4$ shows an example of identification according to the rule \eqref{eqrel}.} Right: The points in $S$ corresponding to  {$C_1, C_2, C_3, C_4, C_5$} in the left diagram, including their causal relations.}
\label{observerdepst}
\end{figure}

We can call $\mathbf{c}_\pm(\sigma_{\theta(t)})$ the future ($+$) or past ($-$) generalized light cone of the set $B_\epsilon(x(t))$ as the projection through $\pi$ of the integral curves of $w \in TE$ starting from ($+$) or ending in ($-$) the set $\pi^{-1}(B_\epsilon(x(t)))$, see Fig. \ref{lightcones}. We use the word \textit{generalized} because, according to Assumption \ref{A2}, particles may appear to move faster than light in $M$. This peculiar fact, crucial for obtaining relative locality, will be discussed in sec. \ref{generalrelloc}. One can then use this structure to introduce a partial order relation $\preceq$ on the elements of $S$ depending on whether one of them is in the future or past generalized light cone of the other or not, see e.g. \cite{Bombelli:1987aa, Surya:2019ndm}. An example is shown in Fig. \ref{observerdepst}. Note that the causal structure identifies points in $S$ which may otherwise differ due to the Rietdijk–Putnam argument discussed in sec. \ref{Secobservers}. In our model, $S$ naturally assumes the structure of a causal set where each element, in addition to causal relations, also possesses intrinsic information (in our case, the sections measured by the observer). Related, though distinct, ideas were introduced in \cite{Cortes:2013uka, Cortes:2013pba}, where causal sets are also enhanced by including intrinsic properties. However, unlike in our model, in \cite{Cortes:2013uka, Cortes:2013pba} these enhanced causal sets replace spacetime, following a line of reasoning similar to that in \cite{Bombelli:1987aa}.

\section{Generalized relative locality}\label{generalrelloc}
We start this section by presenting a simple example.
In order to test whether locality is relative or absolute, Alice and Bob agree beforehand that Bob will send Alice three particles (identical in mass for simplicity) with momenta, e.g., $p_1 > p_2 > p_3$, launching them in the same direction at exactly the same event according to Bob. Alice and Bob also agree on a certain spatial separation $D$ between them. At some point, Bob sends the three particles towards Alice. Once the first particle arrives, Alice (knowing $D$, $p_1, p_2, p_3$) can determine how much time she has to wait before measuring $p_2$ and $p_3$. In absence of deformation, Alice's predictions would be confirmed by her measurements. However, if deformation is present, Alice may see particles arriving at different spacetime points then expected, for example they may arrive at the same time even if $p_1 \neq p_2 \neq p_3$. Since Bob sent the particles at the same event, and since Alice can measure that $p_1, p_2, p_3$ correspond to the agreed-upon values, the only conclusion that she can reach is that it appears as though these particles were launched at different spacetime points. In other words, a single spacetime point for the distant Bob does not correspond to a single spacetime point for Alice, i.e. there is relative locality. 

\begin{figure}[b!]
\centering
\includegraphics[width=0.5\textwidth]{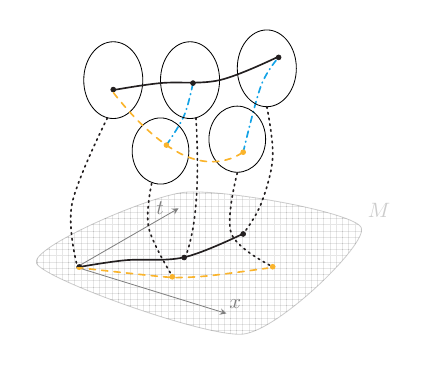} 
\caption{The integral curve of $v$ (upper black line) is displaced to the integral curve of $w = v+\Lambda$ (upper dashed orange line). The azure dotted-dashed lines represent the curves $c_{\sigma(t)\rightarrow f(\sigma(t))}$. In the base manifold, the projections are represented in the same style as the corresponding sections.}
\label{singlepart}
\end{figure}

Assumptions \ref{A1} and \ref{A2} allow us to easily model this effect. For simplicity, we illustrate it by focusing on freely propagating point-like particles (the most common example in the literature, see, e.g., \cite{Amelino-Camelia:2011lvm, Freidel:2011mt, Arzano:2021scz}), and we restrict $M$ to Minkowski spacetime. Nevertheless, both Assumptions \ref{A1} and \ref{A2} hold more generally. A representation of this construction is shown in Fig. \ref{singlepart}.

Consider a vector field $v \in TE$ describing the non-deformed dynamics of a particle. Consider an integral curve $\sigma$ of $v$ starting from some initial condition $e \in E$, the projection on $M$ gives the motion in spacetime of the particle. The pushforward $\pi_*v$ describes the velocity of the particle at a certain point. Notice that, in general, for free particles, momentum and velocity are related by\footnote{We are using this relation because of the restriction to Minkowski spacetime in this example. The argument works in the same way in more general contexts if this relation is substituted for $g(\dot{x}, p, m) = 0$ for some fixed $g$.} $\dot{x} = p/m$. We now add the Ehresmann connection $\Lambda$. The integral curves $\Sigma$ of the new vector field $w = v + \Lambda$ are of course different than those of $v$ (initial conditions are held constant, i.e. the integral curves of $v$ and $w$ start at the same $e\in E$ dictated by the description of the system by a local observer). To visualize the situation, it may be easier to consider a diffeomorphism $f:E \rightarrow E$ mapping $\sigma$ to $f(\sigma) =: \Sigma$. We can now consider the class of integral curves $c_{\sigma(t) \rightarrow f(\sigma(t))}$ of $\Lambda$ alone, starting at $\sigma(t)$ and ending at $f(\sigma(t))$ (for all $t$). We can parallel transport each fibre through $c_{\sigma(t) \rightarrow f(\sigma(t))}$ with the connection $\Lambda$. In this very simplified case, for example, $\Lambda$ can be the trivial horizontal lift from a constant vector field on $M$ (e.g. with $|\Lambda_p| = \lambda \, \,  \forall p \in M$ in order to satisfy Assumption \ref{A2}), so that the fibre coordinate along $c_{\sigma(t) \rightarrow f(\sigma(t))}$ does not change. In other words, we are modifying the spacetime trajectory of the particle but not its momentum, so that now the particle is `off-shell', $\dot{x} \neq p/m$, and can potentially `move faster than light' with respect to the inherent causal structure of the Lorentzian $M$. Such a phenomenon has been very recently shown in $\kappa$-Minkowski spacetime \cite{Kurkov:2025soi} following a different approach.

Recall, however, that any observable consequence must be described in $S$ because of Assumption \ref{A1}. The apparent off-shell velocity of the particle just changes the chain in the causal set describing its evolution\footnote{See also sec. \ref{connection}: a deformation of the light cone in general changes causal relations.}, but observers in $S$ only measure momenta. In our example, Alice expects to measure the particles launched by Bob at specific nodes along her causal path. However, the deformation alters the particles chains, allowing her to detect particles with the same expected momentum but at unexpected nodes.\footnote{It is worth noting that, for high particle densities in the universe, the sampling of events in $S$ corresponds to a high density causal set, which more faithfully reproduces geometric structures in $M$, see e.g. \cite{Bombelli:1987aa, Surya:2019ndm}.} Alice therefore measures relative locality.
It is interesting to notice that, in the example provided above, the linearity of particles motion in $M$ and the choice of a constant $\Lambda$ naturally reproduce a relative locality effect proportional to the distance (in $M$) between Alice and Bob, in agreement with important literature results (see e.g. \cite{Arzano:2021scz} for a review).

We now make a few observations on the nature of relative locality as modeled in this framework.
First, our construction does not require non-trivial interaction vertices, relative locality can arise even when particles depart from the same spacetime point without meaningful interaction. Although interacting particles provide the standard example, interaction is not essential.
Secondly, as shown above, the simplest possible case already reproduces known results from the literature. However, Assumptions \ref{A1} and \ref{A2} are considerably more general, allowing for a definition of relative locality in more general models, including fields. Depending on the properties of the connection $\Lambda$, this framework enables a classification of possible relative locality effects. Additionally, traditional models based on momentum-space curvature can only describe kinematical effects (momentum-space is curved for all particles), while our model allows for spin-dependent connections, describing a different type of relative locality for different types of particles. Moreover, non-trivial topology in the fibre/manifold can lead to a new type of \textit{topological relative locality}\footnote{
The intuitive idea is that, in order to observe topological effects, sections would need to be extended globally. However, such global sections often do not exist (e.g., in non-trivial principal bundles). For any \textit{topological observer} defined through the attempt to extend local sections to global ones, there would therefore be points in the universe to which no momentum could be assigned in the first place, and which would thus be unobservable. These effects could also manifest in the absence of deformation. Additionally, non-trivial holonomy of the connection $\Lambda$ may result in another type of topological relative locality effects. An investigation of these possibilities will be presented in forthcoming papers.}. Finally, our model does not require a curved momentum-space to exhibit relative locality. As shown in sec.~\ref{relloc}, curvature of momentum-space naturally induces a suitable connection $\Lambda$ and therefore implies relative locality, but it is not a necessary condition. To our knowledge, these generalizations, which are naturally included in our model, have not previously appeared in the literature.

\section{Relative locality and momentum-space curvature}\label{relloc}

We have shown that, in our model, relative locality is a consequence of a choice of connection $\Lambda$ depending on the parameters $\lambda$. 
If momentum-space is curved, with curvature parametrized by some coefficients $\lambda$, then, considering the fibre bundle embedded into some $\mathbb{R}^n$ for a suitable $n$ depending on the considered model, the information about the fibre curvature given by the Riemann tensor $R$ is completely contained in its second fundamental form $T(P,Q):= H\nabla_{P}Q$, where $P,Q \in VE$. The relation linking the two quantities is the Gauss curvature formula 
\begin{equation}
    \langle R(X,Y)Z,W \rangle =  \langle T(Y,Z), T(X,W)\rangle
    - \langle T(X,Z), T(Y,W) \rangle
\end{equation}
where $\langle \cdot, \cdot \rangle$ is the inner product on $TE$ \cite{Spivak1979ACI}. 
The second fundamental form defines a connection when evaluated for all $P,Q \in VE$ which, assuming no extrinsic curvature of the fibre embedding, becomes degenerate when $\lambda \rightarrow 0$. In other words, curvature of the fibre is naturally related to a connection needed by Assumption \ref{A2}, which explains the fact that, historically, relative locality has been discovered studying curved spacetimes first.

\section{ {Spacetime symmetries}}\label{symmetries}

 {
A direct consequence of Assumption \ref{A2} is that all spacetime intervals are (locally) modified (with respect to the non-deformed ones) because of the addition of the Ehresmann connection $\Lambda$. This can be described equivalently and intrinsically in $M$ by a map $G: \mathcal{M} \rightarrow \mathcal{M}$ (where $\mathcal{M}$ is the space of metrics on $M$) defined by $g_x \mapsto g_{x, \lambda}$, where $g_{x, \lambda}$ is a $\lambda$-dependent metric and $x \in M$. This map modifies the intrinsic light-cones (and all intervals in general) to make them coincide with those defined by Assumption \ref{A2}. The spacetime symmetries, then, are those which leave $g_{x,\lambda} (v, w)$ unchanged, for any $v, w \in T_xM$ and $x \in M$. In the literature it is common to write $g_{x,\lambda} = e_\lambda g_x e^T_\lambda$ for some appropriate $e_\lambda$, so that a symmetry transformation $R g_{x, \lambda} R^T = g_{x,\lambda}$ takes the form $(e^{-1}_\lambda R e_\lambda) g_x (e^{-1}_\lambda R e_\lambda)^T =: R_\lambda g_x R^T_\lambda = g_x$. In other words, the metric is not affected by deformation, but symmetry generators are deformed by $\lambda$ \cite{Kowalski-Glikman:2002iba, Amelino-Camelia:2000stu, Bevilacqua:2023pqz, Lukierski:2023gxf}. The two approaches are equivalent in our model. Notice that, in cases where momentum space manifold is a Lie group, one often deals not only with symmetry algebras but with symmetry Hopf algebras. However, the group structure is sufficient for the determination of the Hopf algebra properties, see e.g. \cite{Arzano:2022ewc}. Concluding, we note that one can use the imposition of symmetry properties as a tool to identify and study relative locality in different theories. Indeed, given a ($\lambda$-dependent) spacetime symmetry group $\mathcal{G}$, one may reconstruct $g_{x, \lambda}$ (using appropriately defined $e_\lambda$) and from it the vector field $\Lambda$, which in turn allows one to discuss relative locality. This gives a direct link between symmetries and relative locality effects, and may be a powerful tool in the classification of such effects. These interesting research lines will be investigated in forthcoming papers.}

\section{Conclusion and outlook}\label{conclusion}

Building on a fairly straightforward phenomenological definition of observer-dependent spacetime (Assumption \ref{A1}), and on a slight modification of a very general mathematical description of system dynamics (Assumption \ref{A2}), this model allows for the description of relative locality in very general contexts. Using a bundle with spacetime as base manifold is in line with other models \cite{Pfeifer:2021tas} which however lack the natural ability to introduce relative locality in their formalism. Causality is also manifest for general deformed theories in our model, which can therefore be applied to cosmology with $M$ playing the role of cosmological spacetime. The dependence of deformation on the observer is a known result in the literature \cite{Amelino-Camelia:2011lvm, Freidel:2011mt, Carmona:2021gbg}. In deformed theories, the choice of observer also determines the choice of basis in momentum-space, with different bases corresponding to different physical interpretations \cite{Kowalski-Glikman:2002iba, Meljanac:2012pv}. In our model, the choice of observer is linked to the choice of connection. This immediately allows us to determine which observers are equivalent and which are not, by examining the equivalence classes of connections under gauge transformations. The relation between observers and gauges is also important in cosmological models, particularly in the presence of perturbations \cite{Boldrin:2021xrm, Boldrin:2024zmb}, to which this framework could be applied. Moreover, one can also deduce the gauge-independent (i.e. observer-independent) signatures of relative locality, and how to encode them into observables defined in $S$, greatly enhancing our understanding of the physical basis behind PQG models. These interesting projects represent the next natural step of the above model, and will be investigated in future publications.

\section*{Acknowledgements}

We thank Jerzy Kowalski-Glikman for reading a preliminary draft of the paper and for providing useful guidance. This work has benefited from discussions and networking activities carried out within the framework of the COST Actions  {“Bridging High and Low Energies in Search of Quantum Gravity (BridgeQG)”} (CA23130), and  {“Relativistic Quantum Information (RQI)”} (CA23115), supported by COST (European Cooperation in Science and Technology). One of the authors also benefited from discussions with Christian Pfeifer at the School on Quantum Gravity Phenomenology in the Multi-Messenger Approach (COST CA18108 First Training School), Corfu, and from discussions with Anna Pacho\l.  {We also thank the anonimous referee, whose observations greately helped clarify the paper and expand its reach.} This research did not receive any specific grant from funding agencies in the public, commercial, or not-for-profit sectors.

\bibliographystyle{elsarticle-num}

\bibliography{references}

\end{document}